\documentclass[conference]{IEEEtran}
\usepackage{graphicx}
\usepackage{cite}
\usepackage{amsmath}
\usepackage{amsfonts}
\usepackage{amssymb}
\usepackage{subfigure}

\newcommand{\bm}{\mathbf}
\newcommand{\be}{\begin{equation}}
\newcommand{\ee}{\end{equation}}
\newcommand{\bea}{\begin{eqnarray}}
\newcommand{\eea}{\end{eqnarray}}
\newcommand{\x}{{\bm x}}

\newcommand{\q}{{\bm q}}

\newcommand{\bA}{{\bm A}}
\newcommand{\bI}{{\bm I}}

\newcommand{\bW}{{\bm W}}

\newcommand{\bD}{{\bf D}}

\newcommand{\bH}{{\bf H}}

\newcommand{\h}{{\bf h}}

\newcommand{\bv}{{\bf v}}
\newcommand{\bw}{{\bf w}}

\newcommand{\bs}{{\bf s}}

\newcommand{\I}{{\bm I }}

\title{Filter Bank Multicarrier for Massive MIMO}

\author{\normalsize Arman Farhang$^*$, Nicola Marchetti$^*$, Linda E. Doyle$^*$ and Behrouz Farhang-Boroujeny$^\dagger$  
\\$^*$CTVR / The Telecommunications Research Centre, Trinity College Dublin, Ireland, \\
$^\dagger$ECE Department, University of Utah, USA. \\
Email: \{farhanga, marchetn, ledoyle\}@tcd.ie, farhang@ece.utah.edu}

\begin{document}

\maketitle

\begin{abstract}
This paper introduces filter bank multicarrier (FBMC) as a potential candidate in the application of massive MIMO communication. It also points out the advantages of FBMC over OFDM (orthogonal frequency division multiplexing) in the application of massive MIMO. The absence of cyclic prefix in FBMC increases the bandwidth efficiency. In addition, FBMC allows carrier aggregation straightforwardly. Self-equalization, a property of FBMC in massive MIMO that is introduced in this paper, has the impact of reducing (i) complexity; (ii) sensitivity to carrier frequency offset (CFO); (iii) peak-to-average power ratio (PAPR); (iv) system latency; and (v) increasing bandwidth efficiency. The numerical results that corroborate these claims are presented. 
\end{abstract}

\section{Introduction}
\label{sec:intro}
In recent past, massive MIMO has gained significant momentum as a potential candidate to increase the capacity of multiuser networks. In essence, massive MIMO is a multiuser technique (somewhat similar to code division multiple access - CDMA) where the spreading gains for each user are determined by the channel gains between the respective mobile terminal (MT) antenna and the multiple antennas (many of them) at base station (BS). Accordingly, by increasing the number of antennas at BS, the processing gain can be increased to become arbitrarily large. As discussed in the pioneering work of Marzetta \cite{Marzetta2010}, in the limit, as the number of BS antennas tends to infinity, the processing gain of the system tends to infinity and, as a result, the effects of both noise and multiuser interference (MUI) are completely removed. Therefore, the network capacity (in theory) can be increased unboundedly by increasing the number of antennas at the BS, \cite{Marzetta2010}. 

Motivated by Marzetta's observations, \cite{Marzetta2010}, multiple research groups in recent years have studied a variety of implementation issues related to massive MIMO systems, e.g., \cite{Hoydis2013, Hoydis2012, Jose2010, Gao2011, Payami2012}. Also, different groups have started development of testbeds to confirm the theoretical observation of \cite{Marzetta2010}, in practice. An assumption made by Marzetta \cite{Marzetta2010} and followed by other researchers is that orthogonal frequency division multiplexing (OFDM) may be used to convert the frequency selective channels between each MT and the multiple antennas at the BS to a set of flat fading channels. Accordingly, the flat gains associated with the set of channels within each subcarrier constitute the spreading gain vector that is used for dispreading of the respective data stream.    

In this paper, embarking on the above concept, we introduce the application of filter bank multicarrier (FBMC) to the area of massive MIMO communications. FBMC methods have their roots in the pioneering works of Chang \cite{Chang66} and Saltzberg \cite{Saltzberg67}, who introduced multicarrier techniques over two decades before introduction and application of OFDM to wireless communication systems. While OFDM relies on cyclic prefix (CP) samples to avoid intersymbol interference (ISI) and to convert the channel to a set of subcarrier channels with flat gains (perfectly, when channel impulse response duration is shorter than CP), FBMC, without using CP, by adopting a sufficiently large number of subcarriers, relies on the fact that when each subcarrier band is narrow, it is characterized by an approximately flat gain, hence, may suffer only from a negligible level of ISI. 

A new and interesting finding in this paper is that in the case of massive MIMO systems, linear combining of the signal components from different channels smooths channel distortion. Hence, one may relax on the requirement of having approximately flat gain for the subcarriers. This observation, which is confirmed numerically in this paper, positions FBMC as a strong candidate in the application of massive MIMO. As a result, in a massive MIMO setup, one may significantly reduce the number of subcarriers in an FBMC system. This reduces both system complexity and the latency/delay caused by the synthesis filter bank (at the transmitter) and the analysis filter bank (at the receiver). Also, since linear combining of the signal components equalizes the channel gain across each subcarrier, one may adopt larger constellation sizes, hence, further improve on the system bandwidth efficiency. Moreover, increasing the subcarrier spacing has the obvious benefit of reducing the sensitivity to carrier frequency.

An additional benefit of FBMC here is that carrier/spectral aggregation (i.e., using non-contiguous bands of spectrum for transmission) becomes a trivial task, since each subcarrier band is confined to an assigned range and has a negligible interference to other bands. This is not the case in OFDM, \cite{Farhang2011}. 

This paper is organized as follows. A summary of FBMC methods is presented in Section~\ref{sec:FBMC}. From the various choices of FBMC methods, the cosine modulated multitone (CMT) is identified as the best choice in the application of interest in this paper. CMT provides a simple blind tracking of the channel which may prove very beneficial in massive MIMO systems. Hence, to pave the way for new developments in the rest of the paper, a summary of CMT along with its blind equalization capability are presented in Section~\ref{sec:BlindEq}. Section~\ref{sec:CMT} discusses the application of CMT to massive MIMO systems. In this introductory study, we ignore the issues related to channel estimation, including the pilot contamination problem \cite{Marzetta2010}, and simply assume perfect knowledge of channel state information is available to the BS. Self-equalization property of FBMC systems in massive MIMO channels is introduced in Section~\ref{sec:equalization}. Section~\ref{sec:Comparison} presents a qualitative comparison of FBMC and OFDM systems in massive MIMO systems. The numerical results are presented in Section~\ref{sec:NR}. Finally, the conclusions of the paper and suggestions for further research are presented in Section~\ref{sec:Conclusion}.

Throughout this paper, we use the following notations. Scalars are represented in regular upper and lower case letters. Vectors are represented by boldface lower case letters. Matrices are represented by boldface upper case letters. The notation `$\mathbf{I}_N$' denotes an $N\times N$ identity matrix. The matrix or vector superscript $(\cdot)^{\rm T}$ indicates transpose. $\| \cdot \|$ and $|\cdot|$ demonstrate Euclidean norm and absolute value, respectively. Finally, ${\rm diag}(\x)$, $\mathbb{E}[\cdot]$ and $(\cdot)^{-1}$ identify a diagonal matrix with diagonal elements of the vector $\x$, expectation and inverse of a matrix.  

\section{FBMC Methods}\label{sec:FBMC}
The first proposal of the filter bank multicarrier (FBMC) technique came from Chang, \cite{Chang66}, who presented the conditions required for signaling a parallel set of pulse amplitude modulated (PAM) symbol sequences through a bank of overlapping filters. To maximize the bandwidth efficiency of the system, vestigial side-band modulation was applied to each subcarrier signal. Saltzberg, \cite{Saltzberg67}, extended the idea and showed how the Chang's method could be modified for transmission of quadrature amplitude modulated (QAM) symbols. In the literature, this method is often referred to as offset-QAM (OQAM) OFDM. Efficient digital implementation of Saltzberg's multicarrier system through polyphase structures was first studied by Hirosaki \cite{Hirosaki81}, and was further developed by others \cite{Bolcskei03,Bolcskeib03,Hirosaki86,Cariolaro95,Vang01,Siohan02}. 

The pioneering work of Chang, \cite{Chang66}, on the other hand, has received less attention within the signal processing community. Those who have cited \cite{Chang66}, have only acknowledged its existence without presenting much details, e.g., \cite{Hirosaki86} and \cite{Vang01}.  However, the cosine modulated filter banks that have been widely studied within the signal processing community, \cite{VaidBook}, are basically a reinvention of Chang's filter bank, formulated in discrete-time. The use of cosine modulated filter banks for data transmission was first presented in \cite{Tzannes93} and further studied in \cite{Sandberg95}, under the name of discrete wavelet multitone (DWMT). Many other researchers subsequently studied and evaluated DWMT; see \cite{Farhang03} and the references therein. Moreover, \cite{Farhang03} suggested a blind equalization method for cosine modulated based/DWMT systems. An analysis of this blind equalizer was later developed in \cite{LinFar06}. The name cosine modulated multitone (CMT), that we use in the rest of this paper, to refer to the Chang's type of multicarrier modulation was also coined in \cite{LinFar06}. Another relevant work is \cite{FarYuen2010}, where the authors have suggested the shorter name of staggered multitone (SMT) to replace OQAM-OFDM and shown that CMT and SMT are related through a modulation step. 

Both CMT and SMT can be adopted for massive MIMO, leading to the same performance. However, it turns out that derivations and explanation of the results in the context of CMT are easier to follow. We thus limit our attention in the subsequent sections of this paper to development of CMT in massive MIMO application.

\section{CMT and Blind Equalization}\label{sec:BlindEq}
\begin{figure}
\centering
\includegraphics[scale=0.4]{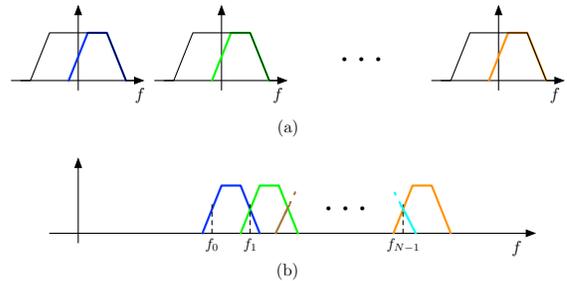}
\vspace{-2 mm}
\caption{CMT modulation. (a) Spectra of baseband data streams (black) and VSB portion of each (other colors). (b) CMT spectrum consisting of modulated versions of the VSB spectra of the baseband data streams. VSB signals are modulated to the subcarrier frequencies $f_0$, $f_1$, $\cdots$, $f_{N-1}$.}
\label{fig:CMTmodulation}
\vspace{-2 mm}
\end{figure}

In CMT a set of pulse amplitude modulated (PAM) baseband data streams are vestigial side-band (VSB) modulated and placed at different subcarriers. Fig.~\ref{fig:CMTmodulation} depicts this process. Moreover, to allow separation of the data symbols (free of ISI and ICI), at the receiver, the carrier phase of the VSB signals is toggled between 0 and $\pi/2$ among adjacent subcarriers. The detailed equations explaining why this approach works can be found in \cite{Chang66} and many other publications; a recommended reference is \cite{FarYuen2010}. Reference \cite{FarhangSDRbook} also provides more details, including the implementation structures and their relevant MATLAB codes.
  
\begin{figure}[t]
\centering
\includegraphics[scale=0.7]{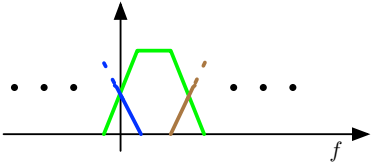}
\vspace{-3 mm}
\caption{Spectrum of a subcarrier in a CMT system after demodulation to baseband, but before matched filtering. Note that interference from the remaining subcarriers is present. Most of such interference will be removed after matched filtering, but still some residual interference from adjacent subcarriers (those at the corners of the transmission band) will remain.}
\label{fig:CMTdemodulation}
\vspace{-3 mm}
\end{figure}

Demodulation of each subcarrier in CMT is a four step procedure.
\begin{enumerate}
\item
For each subcarrier, say, the $k^{\rm th}$ one, the received signal is down-converted to baseband using $f_k$ as the carrier frequency. This results in a spectrum similar to the one presented in Fig.~\ref{fig:CMTdemodulation}.
\item
The demodulated signal is passed through a matched filter that extracts the desired VSB signal at baseband. The matched filter removes most of the signal spectra from other subcarriers. However, as may be understood from Fig.~\ref{fig:CMTdemodulation}, some residuals of adjacent subcarriers remain.
\item
The channel effect is removed from the demodulated signal using a complex-valued single tap equalizer. This is based on the assumption that each subcarrier band is sufficiently narrow such that it can be approximated by a flat gain. A multi-tap equalizer may be adopted if this approximation is not valid. 
\item
After equalization, the real part of VSB signal contains the desired PAM symbol only. Its imaginary part consists of a mix of ISI components and ICI components from the two adjacent bands. Accordingly, taking the real part of the equalized VSB signal delivers the desired data signal/symbol, free of ISI and ICI.    
\end{enumerate}

In \cite{Farhang03},  it is noted that since the ISI and ICI terms are linear contributions from a relatively large number of data symbols, the summation of ISI and ICI has a Gaussian like distribution. This means, while the real part of the equalized signal of each subcarrier channel has contribution from one symbol, hence, has a distribution that follows that of the respective PAM symbols, its imaginary part has a distribution close to that of a Gaussian.
When the channel is unequalized, both real and imaginary parts suffer from both ISI and ICI and in that case, the real part of subcarrier channel output will be a mixture of PAM and a Gaussian distribution. Taking note of this property, it was noted in \cite{Farhang03} that a blind equalizer similar to Godard blind equalization algorithm \cite{Godard80} may be adopted. Such algorithm is developed by minimizing the cost function
\be
\xi=\mathbb{E}[(|y_k(n)|^p-R)^2],
\ee
 where $y_k(n)$ is the equalizer output (in the case here, the equalizer output of the $k^{\rm th}$ subcarrier channel) and $p$ is an integer (usually set equal to 2),
\be
 R=\frac{\mathbb{E}[|s|^{2p}]}{\mathbb{E}[|s|^p]},
\ee
and $s$ is a random selection from the PAM symbols alphabet. A least mean square (LMS) like algorithm is then developed to adjust the single tap equalizers coefficients.

Extension of the above blind equalization method to massive MIMO systems is straightforward. In this paper, we do not pursue this line of research, partly, because of the limited space. However, we wish to point out that application of blind equalization to track channel drifts during the uplink transmission, to have the most updated channel when the BS switches to downlink transmission, may prove very useful in practice. This remains an open research problem for future.

\section{CMT Application to massive MIMO}\label{sec:CMT}
We consider a multiuser MIMO setup similar to the one discussed in \cite{Marzetta2010}. There are $K$ MTs and a BS in a cell. Each MT is equipped with a single transmit and receive antenna, communicating with the cell BS in a {\em time division duplexing} (TDD) manner. The BS is equipped with $N\gg K$ transmit/receive antennas that are used to communicate with the $K$ MTs in the cell {\em simultaneously}. We also assume, similar to \cite{Marzetta2010}, multicarrier modulation is used for data transmissions. However, we replace OFDM modulation by CMT modulation.

Each MT is distinguished by the BS using the respective subcarrier gains between its antenna and the BS antennas. Ignoring the time and subcarrier indices in our formulation, for simplicity of equations, a transmit symbol $s(\ell)$ from the $\ell^{\rm th}$ MT arrives at the BS as a vector 
\be
\x_\ell=(s(\ell)+jq(\ell))\h_\ell,
\ee
where  $\h_\ell=[h_\ell(0),\ldots,h_\ell(N-1)]^{\rm T}$ is the channel gain vector whose elements are the gains between the $\ell^{\rm th}$ MT and different antennas at the BS. $q(\ell$) is the contribution of ISI and ICI. The vector $\x_\ell$ and similar contributions from other MTs, as well as the channel noise vector $\bv$, add up and form the BS received signal vector
\be\label{eqn:x}
\x=\sum_{\ell=0}^{K-1} \x_\ell +\bv.
\ee

The BS uses a set of linear estimators that all take $\x$ as their input and provide the estimates of the users' data symbols $s(0)$, $s(1)$, $\cdots$, $s(K-1)$ at the output. To cast this process in a mathematical formulation and allow introduction of various choices of estimators, we proceed as follows. We define $\tilde\x=[\x_{\rm R}^{\rm T}~\x_{\rm I}^{\rm T}]^{\rm T}$, $\tilde\bv=[\bv_{\rm R}^{\rm T}~\bv_{\rm I}^{\rm T}]^{\rm T}$, $\tilde\h_\ell=[\h_{{\ell,\rm R}}^{\rm T}~\h_{\ell,{\rm I}}^{\rm T}]^{\rm T}$, $\breve\h_\ell=[-\h_{{\ell,\rm I}}^{\rm T}~\h_{\ell,{\rm R}}^{\rm T}]^{\rm T}$,  $\bs=\left[s(0)~s(1)~\cdots~s(K-1)\right]^{\rm T}$ and $\q=\left[q(0)~q(1)~\cdots~q(K-1)\right]^{\rm T}$, where the subscripts `R' and `I' denote the real and imaginary parts, respectively. Using these definitions, (\ref{eqn:x}) may be rearranged as
\be\label{eqn:x3}
\tilde\x=\bA\left[\begin{array}{c}\bs\\ \q\end{array}\right]+\tilde\bv,
\ee
where $\bA=[ \tilde\bH~\breve\bH]$, and $\tilde\bH$ and $\breve\bH$ are $2N\times K$ matrices with columns of $\{\tilde\h_\ell,~\ell=0,1,\cdots,K-1\}$ and $\{\breve\h_\ell,~\ell=0,1,\cdots,K-1\}$, respectively. Equation (\ref{eqn:x3}) has the familiar form that appears in CDMA literature, e.g., see \cite{Madhow94,FarhangBook2013}. Hence, a variety of solutions that have been given for CDMA systems can be immediately applied to the present problem as well. For instance, the matched filter (MF) detector obtains an estimate of the vector $\bs$, according to the equation
\be\label{eqn:MF}
\hat{\bs}_{\rm MF}=\bD^{-1}{\boldsymbol \Gamma}\bA^{\rm T}\tilde\x,
\ee 
where $\bD={\rm diag}\{\|{\tilde\h_0}\|^2,\ldots,\|{\tilde\h_{K-1}}\|^2\}$, the matrix ${\boldsymbol \Gamma}$ consists of the first $K$ rows of the identity matrix $\bI_{2K}$ and $\hat{\bs}_{\rm MF} = [\hat{s}_{\rm MF}(0),\ldots,\hat{s}_{\rm MF}(K-1)]^{\rm T}$ whose $\ell^{\rm th}$ element, $\hat{s}_{\rm MF}{(\ell)}$, is the estimated data symbol of user $\ell$. Using (\ref{eqn:MF}), each element of $\hat{\bs}_{\rm MF}$ can be expanded as
\be\label{eqn:MFl}
\hat{s}_{\rm MF}{(\ell)}=s(\ell)+\sum\limits_{\substack{i=0 \\ i \neq \ell}}^{K-1}\frac{\tilde\h_\ell^{\rm T}}{\|{\tilde\h_\ell}\|^2}(\tilde\h_i s(i)+\breve\h_i q(i))+\frac{\tilde\h_\ell^{\rm T}}{\|\tilde\h_\ell\|^2}\tilde\bv.
\ee
This leads to a receiver structure similar to that of \cite{Marzetta2010}, where it is shown that when the number of antennas, $N$, increases to infinity, the multiuser interference and noise effects vanish to zero. Hence, $\hat \bs=\bs$, where the vector $\hat \bs$ is an estimate of $\bs$, and the receiver will be optimum. In the context of CDMA literature, this has the explanation that as $N$ tends to infinity, the processing gain also goes to infinity and accordingly multiuser interference and noise effects vanish.

In realistic situations when $N$ is finite, the MF estimator is not optimal. A superior estimator is the linear minimum mean square error (MMSE) estimator
\be \label{eqn:hats}
\hat \bs=\bW^{\rm T}\tilde\x,
\ee
where the coefficient matrix  $\bW$ is chosen to minimize the cost function
\be\label{eqn:zeta}
\zeta=\mathbb{E}[\|\bs-\bW^{\rm T}\tilde\x\|^2].
\ee
This solution is optimal in the sense that it maximizes the signal-to-interference-plus-noise ratio (SINR), \cite{FarhangBook2013}. 

Following the standard derivations, the optimal choice of $\bW$ is obtained as
\be\label{eqn:Wo}
\bW_{\rm o}=\bA\left(\bA^{\rm T}\bA+{\sigma_v^2}\I_{2K}\right)^{-1}{\boldsymbol \Gamma}^{\rm T}.
\ee
Here, it is assumed that the elements of the noise vector $\tilde\bv$ are independent and identically distributed Gaussian random variables with variances of $\sigma_v^2$, hence, $\mathbb{E}\left[\tilde\bv\tilde\bv^{\rm T}\right]=\sigma_v^2\I$. The columns of $\bW_{\rm o}$ contain the optimal filter tap weights for different users. 

Substitution of (\ref{eqn:Wo}) into (\ref{eqn:hats}) leads to the MMSE solution
\begin{eqnarray}\label{eqn:MMSEl}
\hat{s}_{\rm MMSE}{(\ell)}&=&\bw_{{\rm o},\ell}^{\rm T}{\tilde\h_\ell}s(\ell)+\sum\limits_{\substack{i=0 \\ i \neq \ell}}^{K-1}\bw_{{\rm o},\ell}^{\rm T}{\tilde\h_i}s(i) \nonumber\\
&& +\sum\limits_{i=0}^{K-1}\bw_{{\rm o},\ell}^{\rm T}{\breve\h_i}q(i)+\bw_{{\rm o},\ell}^{\rm T}\tilde\bv, 
\end{eqnarray} 
where $\bw_{{\rm o},\ell}$ is the $\ell^{th}$ column of $\bW_{\rm o}$. Ignoring the off-diagonal elements of $\left(\bA^{\rm T}\bA+\sigma_v^2\I_{2K}\right)$ and also removing the term $\sigma_v^2\I_{2K}$ from (\ref{eqn:Wo}), one will realize that  (\ref{eqn:MMSEl}) boils down to the MF tap weights (\ref{eqn:MFl}).

The first terms on the right hand side of equations (\ref{eqn:MFl}) and (\ref{eqn:MMSEl}) are the desired signal and the rest are the interference plus noise terms. We consider $s(\ell)$ and $q(\ell)$ as independent variables with variance of unity. With the assumption of having a flat channel impulse response in each subcarrier band, SINR at the output of the MF and MMSE detectors for user $\ell$ in a certain subcarrier can be derived, respectively, as

{\small{
\be
\label{eqn:SINR_MF}
{\rm SINR_{MF}}{(\ell)}=\frac{\|{\tilde\h_\ell}\|^4}{\sum\limits_{\substack{i=0 \\ i \neq \ell}}^{K-1}\big({|{\tilde\h_\ell}^{\rm T}{\tilde\h_i}|}^2+{|{\tilde\h_\ell}^{\rm T}{\breve\h_i}|}^2\big)+\sigma_v^2\|{\tilde\h_\ell}\|^2},
\ee}}
and
{\small{
\be
\label{eqn:SINR_MMSE}
{\rm SINR_{\rm MMSE}}{(\ell)}=\frac{|\bw_{{\rm o},\ell}^{\rm T}{\tilde\h_\ell}|^2}{\sum\limits_{\substack{i=0 \\ i \neq \ell}}^{K-1}{|\bw_{{\rm o},\ell}^{\rm T}{\tilde\h_i}|^2}+\sum\limits_{i=0}^{K-1}{|\bw_{{\rm o},\ell}^{\rm T}{\breve\h_i}|^2}+{\sigma_v^2}\|\bw_{{\rm o},\ell}\|^2}.
%\vspace{-5 mm}
\ee}}

\section{Self-Equalization}\label{sec:equalization} 
In the conventional (single-input single-output) FBMC systems, in order to reduce channel equalization to single tap per subcarrier, it is often assumed that the number of subcarriers is very large. Hence, each subcarrier band may be approximated by a flat gain. This, clearly, has the undesirable effect of reducing the symbol rate (per subcarrier) which along with it brings (i) the need for longer pilot preambles (equivalently, reduces the bandwidth efficiency); (ii) increases latency in the channel; (iii) higher sensitivity to carrier frequency offset (CFO); and (iv) higher peak-to-average power ratio (PAPR) due to the large number of subcarriers which increases the dynamic range of the FBMC signal.

Massive MIMO channels have an interesting property that allows us to resolve the above problems. The MF and MMSE detectors that are used to combine signals from the receive antennas average distortions from different channels and thus, as the number of BS antennas increases, result in a nearly equalized gain across each subcarrier band. This property of massive MIMO channels, that we call {\em self-equalization} is confirmed numerically in Section~\ref{sec:NR}. Theoretical evaluation of this property is left for future study.

\section{Comparison with OFDM} \label{sec:Comparison}
In the case of OFDM, the multiuser equation (\ref{eqn:x3}) simplifies to
\be\label{eqn:xOFDM}
\x=\bH\bs+\bv.
\ee 
Here, $\x$ is the vector of the received signal samples (over a specified subcarrier), $\bH$ is the matrix of channel gains, $\bs$ is the vector of data symbols from different users, and $\bv$ is the channel noise vector. All these quantities are complex-valued. 

The following differences pertain if one compares (\ref{eqn:x3}) and (\ref{eqn:xOFDM}). 
\begin{enumerate}
\item
While all variables/constants in (\ref{eqn:x3}) are real-valued, their counterparts in (\ref{eqn:xOFDM}) are complex-valued.
\item
The users' data vector $\bs$ in (\ref{eqn:xOFDM}) has $K$ elements. This means each user receives multiuser interference from $K-1$ other users. In (\ref{eqn:x3}), on the other hand, each user receives interference from $2(K-1)$ users, from which $K-1$ are actual users  and the rest we refer to as virtual users. For instance, if the data user of interest is $s(0)$, it may receive interference from $s(1)$, $s(2)$, $\cdots$, $s(K-1)$ (the actual user symbols) and $q(1)$, $q(2)$, $\cdots$, $q({K-1})$  (the virtual user symbols - contributions from ISI and ICI components). 
\item
While the processing gain in the OFDM-based systems is $N$ (equal to the number of elements is each column of channel gain matrix $\bH$), this number doubles in the CMT-based system. 
\item 
Considering the observations 2) and 3), it is readily concluded that both CMT-based and OFDM-based systems suffer from the same level of multiuser interference.
\end{enumerate}

These observations imply that, in massive MIMO, signal enhancement through linear combining leads to the same results for both OFDM and CMT-based systems. Nevertheless, CMT offers the following advantages over OFDM.

\vspace{1mm}

\noindent{\em More flexible carrier aggregation}: To make a better use of the available spectrum, recent wireless standards put a lot of emphasis on carrier aggregation. Variety of implementations of carrier aggregation has been reported. Apparently, some companies use multiple radios to transmit/receive signals over different portions of the spectrums. Others, e.g., \cite{GFDM}, suggest modulation and filtering of the aggregated spectra. These solutions are more expensive and less flexible than carrier aggregation in FBMC. Hence, one compelling reason to adopt FBMC in future standards may be this advantage that it has over OFDM.

\vspace{1mm}

\noindent{\em Lower sensitivity to CFO}: As mentioned in the previous sections and demonstrated numerically in the next section, compared to OFDM, FBMC allows an increase in subcarrier spacing. This, in turn, reduces the sensitivity of FBMC to CFO.

\vspace{1mm}

\noindent{\em Lower PAPR}: Reduced number of subcarriers naturally brings low PAPR property to the FBMC signal.

\vspace{1mm}

\noindent{\em Higher bandwidth efficiency}: Because of the absence of CP in FBMC, it expectedly offers higher bandwidth efficiency than OFDM. One point to be noted here is that FBMC usually requires a longer preamble than OFDM. The possibility of reducing the number of subcarriers in FBMC that was noted in Section~\ref{sec:equalization} can significantly reduce the preamble length in FBMC. Hence, it reduces the overhead of the preamble to a negligible amount.

\section{Numerical Results}\label{sec:NR}
In this section, a broad set of numerical results is presented to confirm the theoretical developments of this paper. It was noted in the previous sections that massive MIMO, through use of a large number of antennas at the BS, can provide a large processing gain. Hence, noise and MUI effects can be reduced. In addition, when FBMC is used for signal modulation, linear combining of the signals from multiple antennas at the BS has a flattening effect (i.e., the channel will be equalized) over each subcarrier band. This interesting impact of massive MIMO allows reducing the number of subcarriers in FBMC and this, in turn, has the effect of reducing (i) complexity; (ii) the preamble (training) length, hence, increasing bandwidth efficiency; (iii) sensitivity to CFO; (iv) PAPR; and (v) system latency. The first set of results that we present in this section provide evidence of the self-equalization property.
%The numerical results in this section begin with a study of this property in FBMC systems.

\begin{figure}[ht]
		\centering 
		\subfigure[]{\hspace{-9 mm}
    \includegraphics[scale=0.7]{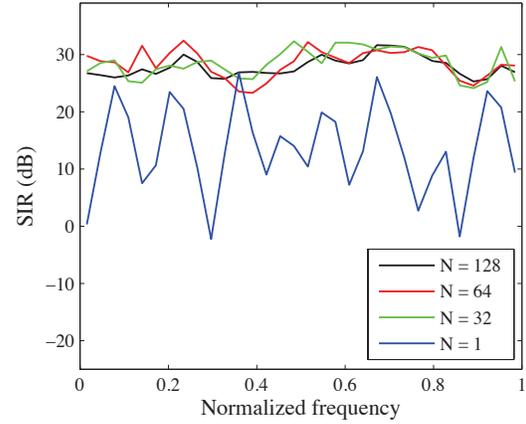}
    \label{fig:L16}
    }
	\subfigure[]{\hspace{-7 mm}
    \includegraphics[scale=0.7]{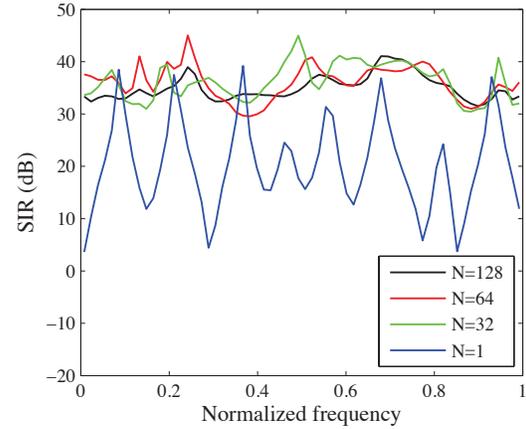}
    \label{fig:L32}
	}
	\subfigure[]{\hspace{-7 mm}
    \includegraphics[scale=0.7]{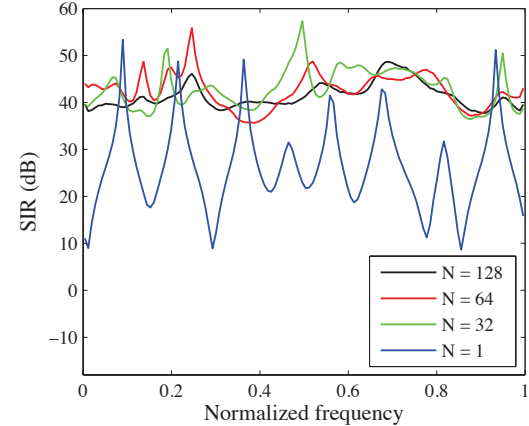}
    \label{fig:L64}
	}
\caption{\subref{fig:L16}, \subref{fig:L32} and \subref{fig:L64} compare the SIR of the MF linear combining technique for the cases of $32,64$ and $128$ subcarriers, respectively, for different number of receive antennas, $N$.}
\vspace{-3.8 mm}
\label{fig:SIR}
\end{figure} 
Fig.~\ref{fig:SIR} presents a set of results that highlights the effect of increasing the number of antennas at the receiver on the signal-to-interference ratio (SIR) for different number of subcarriers in the single-user case. The results are presented for a noise free channel to explore the impact of the number of subcarriers (equivalently, the width of each subcarrier band) and the number of BS antennas in achieving a flat channel response over each subcarrier band. The results are for a sample set of channel responses generated according to the SUI-4 channel model proposed by the IEEE802.16 broadband wireless access working group, \cite{SUI}. SIRs are evaluated at all subcarrier channels. Note that in each curve, the number of points along the normalized frequency is equal to the number of subcarrier bands, $L$. For the channel model used here, the total bandwidth, equivalent to the normalized frequency one, is equal to 2.8~MHz. This, in turn, means the subcarrier spacing in each case is equal to $2800/L$~kHz. As an example, when $L=64$, subcarrier spacing is equal to 87.5~kHz. This, compared to the subcarrier spacing in OFDM-based standards (e.g., IEEE 802.16 and LTE), is relatively broad;  $87.5/15\approx 6$ times larger. Reducing the number of subcarriers (equivalently, increasing symbol rate in each subcarrier band), as noted in Section~\ref{sec:equalization}, reduces transmission latency, increases bandwidth efficiency, and reduces sensitivity to CFO and PAPR. 

\begin{figure}[t]
\centering
%\vspace{-3.8 mm}
\includegraphics[scale=0.7]{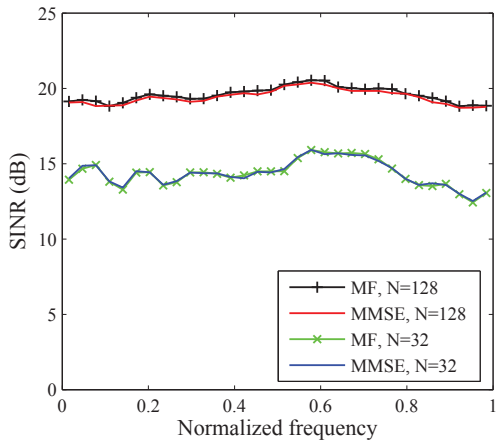}
%\vspace{-1.7 mm}
\caption{SINR comparison between MMSE and MF linear combining techniques in the single user case with $L=32$, when the user's SNR at the receiver input is $-1$ dB for two cases of $N=128$ and $N=32$.}
\label{fig:SINR_SingleUser}
\vspace{-4 mm}
\end{figure}

Next, the channel noise is added to explore similar results to those in Fig.~\ref{fig:SIR}. Since a MF/MMSE receiver in the uplink (or a precoding in the downlink) has a processing gain of $N$, the SINR at the output may be calculated as ${\rm SNR}_{\rm in}+10\log_{10}N$, where ${\rm SNR}_{\rm in}$ is signal-to-noise ratio (SNR) at each BS antenna. The results, presented in Fig.~\ref{fig:SINR_SingleUser}, are for the cases where there are $32$ and $128$ antennas at the BS, the number of subcarriers is equal to $32$ and ${\rm SNR}_{\rm in}=-1$~dB. As seen, here, the SINR curves in both MMSE and MF receivers coincide. The processing gains for $L=128$ and $32$ antennas are respectively $21$ and $15$ dB, and the expected output SINR values $20$ and $14$~dB are observed.

The above results were presented for the single user case. The situation changes significantly in the multiuser scenario due to the presence of MUI. As shown in the following results, when multiple MTs simultaneously communicate with a BS, MMSE outperforms MF by a significant margin. This result that is applicable to both FBMC and OFDM-based MIMO systems is indeed a very interesting observation that has also been recently reported in \cite{Krishnan2013_Nicola}. 

\begin{figure}[t]
\centering
%\vspace{0.5 mm}
\includegraphics[scale=0.7]{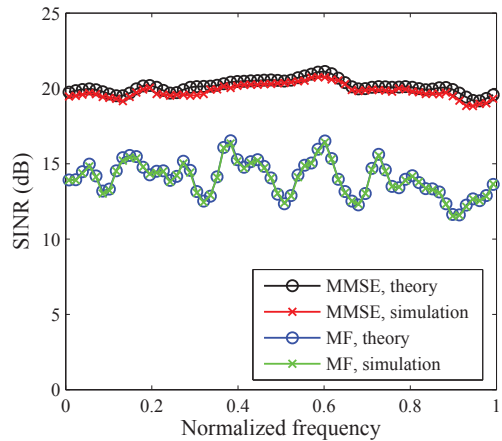}
%\vspace{-0.5 mm}
\caption{SINR comparison between MMSE and MF linear combining techniques for the case of $K=6$, $L=64$ and $N=128$.}
\label{fig:SINR_MultiuserL32}
\end{figure}

\begin{figure}[t]
\centering
\includegraphics[scale=0.7]{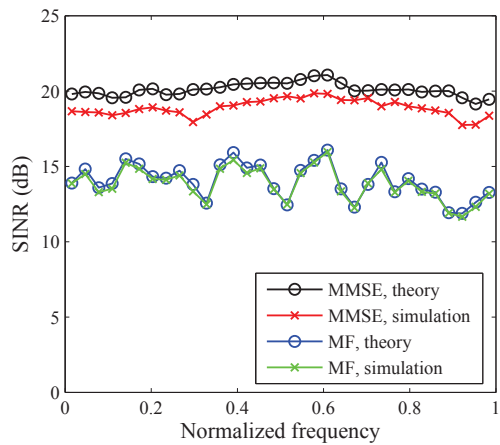}
\vspace{-2 mm}
\caption{SINR comparison between MMSE and MF linear combining techniques for the case of $K=6$, $L=32$ and $N=128$.}
\label{fig:SINR_MultiuserL16}
\vspace{-3 mm}
\end{figure}

The analytical SINR relationships derived in Section~\ref{sec:CMT} are calculated with the assumption of having a flat channel per subcarrier. Therefore, they can be chosen as benchmarks to evaluate the channel flatness in the subcarrier bands.  Figs. \ref{fig:SINR_MultiuserL32} and \ref{fig:SINR_MultiuserL16} present the theoretical and simulations results in a multiuser scenario where $K=6$, $N=128$, the target SINR is $20$ dB (the SNR at each antenna at the BS is selected as $20-10\log_{10} N$ dB) and the cases of $L=64$ and $32$ are examined. As the figures depict, the MMSE combining technique is superior to the MF one and its SINR is about the same for all the subcarriers, i.e., has smaller variance across the subcarriers. When $L=64$ (Fig.~\ref{fig:SINR_MultiuserL32}), the SINR curves for both MF and MMSE techniques, the simulation results coincide with the theoretical ones almost perfectly, confirming the self-equalization property of linear combining in massive MIMO FBMC systems. When $L=32$ (wide-band subcarriers with 87.5 kHz width), the SINR curves from simulations for the MF receiver is still the same as the theoretical curve. However, the SINR simulations for the MMSE combining falls $1$ dB below the theoretical predictions. 

\section{Conclusion}\label{sec:Conclusion}
In this paper, we introduced filter bank multicarrier (FBMC) as a viable candidate in the application of massive MIMO. Among various FBMC techniques, the cosine modulated multitone (CMT) was identified as the best choice. It was shown that while FBMC offers the same processing gain as OFDM, it offers the advantages of more flexible carrier aggregation, higher bandwidth efficiency (because of the absence of CP), blind channel equalization and larger subcarrier spacing, hence, less sensitivity to CFO and lower PAPR. The self-equalization property of CMT in massive MIMO channels was also elaborated on. The SINR performance for two different linear combining techniques; namely, matched filter (MF) and minimum mean squared error (MMSE) linear combiners were investigated. The analytical SINR equations for the aforementioned techniques were derived and their accuracy were evaluated through numerical examples. 

The work presented in this paper initiates a new line of research which may be pursued in a number of directions. Among these, the following research topics may be named at this time. (i) Pilot contamination effect. (ii) Deeper/mathematical study of self-equalization. (iii) Blind channel tracking methods. (iv) CFO analysis. 

\bibliographystyle{IEEEtran}

\end{document}